\documentstyle[preprint,aps]{revtex}
\begin{document}
\draft

\widetext
\title{ Some exact results for the multicomponent $t$-$J$ model }
\author {  Antimo Angelucci$^1$ and  Sandro Sorella$^2$ }
\address{ $^1$ International Centre for Theoretical Physics,
P.O. Box 586, I-34100 Trieste, Italy  \\
$^2$ INFM and International School for Advanced Studies,
Via Beirut 2-4, I--34013 Trieste, Italy}
\date{\today, SISSA preprint: 70/96/CM/MB}
\maketitle
\begin{abstract}
We present a generalization of the Sutherland's multicomponent 
model. Our extension includes both the ferromagnetic and the 
antiferromagnetic $t$-$J$ model for any value of the exchange coupling 
$J$ and the hopping parameter $t$. We prove rigorously that for one dimensional 
chains the ground-state of the generalized model is non-degenerate. As a 
consequence, the ordering of energy levels of the 
antiferromagnetic $t$-$J$ model is determined. Our result rigorously proves 
and extends the analysis carried out by Sutherland in establishing the phase 
diagram of the model as a function of the number of components.

\end{abstract}
\pacs{ 74.20.-z, 71.20.Ad, 75.10.Jm  }


\narrowtext

The $t$-$J$ Hamiltonian is certainly one of the most attractive model 
to explain high temperature superconductivity (HTc) 
without any phonon pairing mechanism\cite{putikka}. Despite its 
simplicity and the huge amount of literature presented starting soon 
after the discovery of HTc, the determination of the full phase 
diagram, especially in two dimensions, is still controversial.

Recently, after a work of Bares {\it et al.}\cite{bares}, it was immediately 
realized that the $t$-$J$ model at the supersymmetric points $J=\pm 2\,t$ 
is a particular case of a multicomponent model introduced previously by 
Sutherland\cite{suth} in his investigation of models solvable 
by Bethe Ansatz. 
In the present work we generalize this mapping as to include the general 
and physically more interesting $J\ne \pm 2\,t $ case. Our approach allows the  
determination of some rigorous and novel result about the $t$-$J$ Hamiltonian 
and other related models. We also discuss in some 
detail the relation between the multicomponent model and the usual form 
of the $t$-$J$ Hamiltonian, trying to clarify some delicate point 
concerning the interpretation of Sutherland's work and 
his notations. 

Consider a hypercubic lattice of $N$ sites and place $N$ objects, or 
particles, on it exactly one to a site (hard-core condition). The 
particles are of $z$ different species, of which $x$ are assumed 
bosonic and $y$ fermionic. We define the multicomponent 
$t$-$J$ model 
\begin{equation}
H = - t \sum_{<ij>} P^K_{i,j} - 
{J\over 2} \sum_{<ij>} P^M_{i,j}, 
\label{tJ}
\end{equation}
where the sum is extended to nearest-neighbor (NN) sites, 
and where $P^K_{i,j}$ and $P^M_{i,j}$ permute particles occupying 
sites $i,j$ if these have opposite or same statistics, 
respectively, otherwise they give zero. The operators in 
Eq.~(\ref{tJ}) are thus defined via the decomposition 
$P_{i,j} = P^K_{i,j}+P^M_{i,j}$ of the  operator $P_{i,j}$ permuting 
whatever pair of particles occupies sites $i,j$, so that for $J=2\,t$ 
Eq.~(\ref{tJ}) reduces -- apart for an overall constant -- to the Sutherland's 
multicomponent Hamiltonian $H_S = - \sum_{<ij>} P_{i,j}$.
We notice that $H$ is equivalent to $H_S$ 
also for the case of a purely bosonic ($y=0$) or fermionic ($x=0$) 
system, as for in this case $P^K_{i,j}\equiv 0$.
The Hamiltonian (\ref{tJ}) is better interpreted as a model for a single 
boson $B$ and a single fermion $F$ having 
$x$ and $y$ different {\em colour} states, respectively. According to this 
simple interpretation, and following the notation of Sutherland, 
we refer to $H_S$ as the $B^xF^y$ system and similarly  
we refer to Eq.~(\ref{tJ}) as the $B^xF^y_J$ system. 
In the latter case we employ the subscript $J$ to stress that
the Hamiltonian (\ref{tJ}) carries a free parameter (i.e., the ratio $J/t$) 
not present in the supersymmetric model $H_S$. 
Henceforth we consider only bipartite lattices. In this case the sign
of the hopping parameter in Eq. (1) is irrelevant, because
on a bipartite lattice it can always be changed by means of a 
unitary transformation\cite{lieb-wu}. Henceforth, we assume $t\ge0$. 
More interestingly, also the sign of the coupling $J$ can be fixed 
without loss of generality, because {\em the $B^xF^y_J$ and $B^yF^x_{-J}$ 
systems are equivalent}\cite{suth}. Henceforth in Eq.~(\ref{tJ}) 
we set $J\ge0$. 
Using this sign {\em convention} the $BF^2_J$ and $B^2F_J$ systems are 
(apart from a shift in the zero of the energy) the usual $t$-$J$ 
Hamiltonian for $J>0$ and $J<0$, 
respectively, whereas the $F^2$ and $B^2$ systems correspond to the 
antiferromagnetic and ferromagnetic Heisenberg models. 
The $BF$ system is instead a simple free spinless fermion Hamiltonian. 
One can also easily check that the $B^2F^2$ system is 
the EKS model introduced by E{\ss}ler {\em et al.}\cite{EKS} in their 
investigation of $\eta$-pairing superconductivity, and that
the $B^2F^2_{J=0}$ system corresponds to a bond-charge Hubbard model as 
discussed by Arrachea and Aligia\cite{aligia}. 

To avoid a too cumbersome notation, we employ a single colour index  
$\alpha=1,2,..,y,y+1,..,z$, with 
first $y$ and last $x=z-y$ labels referring to fermion
and boson colours, respectively. To distinguish the statistics we then 
set $\xi_{\alpha}=1$ $(-1)$ for bosons (fermions).
Henceforth we denote $N_F = \sum_{\alpha=1}^y N^{\alpha}$ and 
$N_B = \sum_{\alpha=y+1}^z N^{\alpha}$ the total number of fermions and
of bosons, respectively, where $N^{\alpha}$ is the number of particles  
of colour $\alpha$. 
We remark that  each number $N^{\alpha}$ is individually conserved
because in Eq.~(\ref{tJ}) there only enter permutators. 
An immediate and important consequence which will be used later is that:
\begin{description}
\item[(A)]
 Each $B^n F^m_J$ system with $n\le x$ and/or $m\le y$ is a particular 
subspace of the $B^x F^y_J$ system Hilbert space, 
\end{description}
which is obtained by setting to zero the occupation numbers of $(x-n)$ 
boson and $(y-m)$ fermion colours.

To describe the model (\ref{tJ}) in second quantization 
we employ a simple oscillator representation where 
$f^{\dagger}_{\alpha,i}$ ($f_{\alpha,i}$) creates (destroys) 
a particle of colour $\alpha$ on site $i$ out of a unphysical reference 
vacuum $\vert 0 \rangle$, and accordingly $f^{\dagger}_{\alpha,i}$, 
$f_{\alpha,i}$ are assumed as fermion (boson) operators for 
$\xi_{\alpha}=-1$ ($\xi_{\alpha}=1$). The hard-core condition 
leads to the set of local constraints $\sum_{\alpha=1}^z 
f^{\dagger}_{\alpha,i}f_{\alpha,i} = 1$, $\forall i$, so that 
a vector of the physical Fock space is given by a product 
of $N$ distinct creation 
operators acting on the vacuum. On any of such Fock vectors the operator 
$P_{i,j}$ acts by interchanging the site indices 
\begin{equation}
P_{i,j} ...f^{\dagger}_{\alpha,i}...f^{\dagger}_{\beta,j} ...\vert 0 \rangle
= ...f^{\dagger}_{\alpha,j}...f^{\dagger}_{\beta,i}...\vert 0 \rangle,
\label{majorana}
\end{equation}
(i.e., it acts like a ``Majorana'' permutator). 
In the physical Fock space we thus have 
\begin{eqnarray} 
P^K_{i,j} & = & \sum_{\{\alpha,\beta\vert\xi_{\alpha}\ne\xi_{\beta}\}}
f^{\dagger}_{\alpha,j}f^{\dagger}_{\beta,i}
f_{\beta,j}f_{\alpha,i}
\nonumber \\ 
P^M_{i,j} & = & \sum_{\{\alpha,\beta\vert\xi_{\alpha}=\xi_{\beta}\}}
f^{\dagger}_{\alpha,j}f^{\dagger}_{\beta,i}
f_{\beta,j}f_{\alpha,i}
\label{permM}
\end{eqnarray} 
We notice that if one forces to represent $P_{i,j}$ by means of 
the Dirac permutator  $P^D_{i,j}$ 
interchanging the colour labels (i.e., $P^D_{i,j} 
f^{\dagger}_{\alpha,i}f^{\dagger}_{\beta,j} \vert 0 \rangle = 
f^{\dagger}_{\beta,i} f^{\dagger}_{\alpha,j}\vert 0 \rangle$), 
then an extra minus sign is involved for the exchanges of two 
fermions (i.e., $P_{i,j} = - P^D_{i,j}$ when permuting two fermions, 
and $P_{i,j}=P^D_{i,j}$ otherwise). Due to  this property, $P_{i,j}$ is 
also defined in literature as a graded permutator\cite{bares,EKS}.
Henceforth we always refer to Eq.~(\ref{majorana}), because 
{\em  this is the definition consistent with 
Sutherland's notations}. 

In the oscillator representation Hubbard operators 
have the simple bilinear  expression 
$\Gamma^{\alpha}_{i,\beta} = f^{\dagger}_{\alpha,i}f_{\beta,i}$. 
As is well known\cite{Wieg}, the total operators 
${\Gamma^{\alpha}}_{\beta} = \sum_{i=1}^N \Gamma^{\alpha}_{i,\beta}$ 
for both fermionic ($\xi_{\alpha} = \xi_{\beta}=-1$) and 
 bosonic ($\xi_{\alpha} = \xi_{\beta}=1$) labels are the 
generators of two commuting SU$(y)$ and SU$(x)$ algebras, respectively. 
The diagonal operators of the algebras are the numbers $N^{\alpha}$, i.e., 
${\Gamma^{\alpha}}_{\alpha} = N^{\alpha}$, whereas the off-diagonal operators 
${\Gamma^{\alpha}}_{\beta}$, for $\alpha\ne \beta$, represent Cartan 
raising-lowering step operators, as for they  change by $\pm 1$ the 
numbers $N^{\alpha},N^{\beta}$.  As  $H$ commutes with both the algebras, 
\begin{equation} 
[H,{\Gamma^{\alpha}}_{\beta}] = 0\quad 
{\rm for}~~\xi_{\alpha} = \xi_{\beta},
\label{comm}
\end{equation}
it follows that for each fixed value of $N_F$ (hence, of 
$N_B=N-N_F$) the spectrum of 
$H$ can be classified in multiplets of SU$(y)$$\otimes$SU$(x)$\cite{susy}. 
Our task is to characterize the sector of the ground-state and, more in 
general, the ordering of the energy levels of the model (\ref{tJ}) in one 
dimension. To this aim, we impose open boundary conditions, 
because this choice greatly simplifies the discussion. 

$\bullet$ We consider the sector $\{N^{\alpha}\}$ with definite values of 
the numbers $N^{\alpha}$. The main result of this paper is the proof that:
\begin{description}
\item[(B)]
{\em  For non-vanishing $t$ and $J$, in any of the possible sectors 
$\{N^{\alpha}\}$ the ground-state of the Hamiltonian (\ref{tJ}) on an 
open chain is non-degenerate}. 
\end{description}

Denoting $i_1,..,i_{N^1}$ the positions of the particles of colour
$\alpha=1$, then $i_{N^1+1},..,i_{N^1+N^2}$ those of the particles of 
colour $\alpha=2$, and so on, a basis $\{\vert a\rangle\}$ of the sector 
is given the $N!/(\prod_{\alpha=1}^z N^{\alpha}!)$ 
distinct sequences  
\begin{equation}
\vert a\rangle = f^{\dagger}_{1,i_1}...f^{\dagger}_{1,i_{N^1}}...
f^{\dagger}_{z,i_{N-N^z+1}}...f^{\dagger}_{z,i_N}\vert 0 \rangle,
\label{state}
\end{equation}
that we order according to the fundamental regions 
\begin{equation}
R_a = \left\{
\begin{array}{ccccccc}
i_1 & < & i_2 & < & ... & < & 
i_{N^1} \\
i_{N^1+1} & < & i_{N^1+2} & < & ... & < &
i_{N^1+N^2} \\
 ... &  ... & ... & ... & ... & ... & ... \\ 
i_{N-N^z+1}  & < & 
i_{N-1}  & < & ... & < & 
i_{N} 
\end{array}\right. ,
\label{region}
\end{equation}
where all operators of same colour are grouped together, and where colour
labels and, separately, coordinates of particles of same colour are ordered 
in ascending order. Because of the open boundary conditions, one can easily 
see that in the chosen basis the  off-diagonal matrix elements 
$H_{a,b}=\langle a\vert H\vert b\rangle$, $a\ne b$, of the Hamiltonian 
(\ref{tJ}) are non-positive. 
From this property, by evaluating the expectation value of the energy over a 
normalized state-vector $\vert \psi \rangle = \sum_a \gamma_a \vert a\rangle$ 
one obtains\cite{lieb-mattis} 
\begin{equation} \label{perron} 
E=\sum\limits_{a,b} H_{a,b} \gamma_a^* \gamma_b \ge 
\sum\limits_{a,b}  H_{a,b} \vert \gamma_a\vert\vert\gamma_b\vert,
\end{equation}
so that a ground-state can be chosen with  non-negative
coefficients, 
$\gamma_a= \vert\gamma_a\vert$.

The non-degeneracy of the ground-state is then proved by showing that for the 
ground-state the coefficient of each element of the basis is non-zero, 
$\gamma_a=\vert\gamma_a\vert\ne 0$, $\forall a$, as for no other 
ground-state with the same restrictions on the $\gamma_a$'s can satisfy the 
orthogonality requirement. According to Perron-Frobenius theorem,
the above more restrictive result holds provided two arbitrary states 
$\vert a \rangle$, $\vert b \rangle$ are linked by the 
Hamiltonian\cite{tian}, namely one can find an integer $Q$ such that
$\langle b \vert H^Q \vert a\rangle\ne0$.
The Hamiltonian $H$ satisfies this property whenever $t$ and $J$ are both 
different from zero. In fact, the site indices $\{ j_k \}$ of the state 
$\vert b\rangle = f^{\dagger}_{1,j_1}...f^{\dagger}_{z,j_N}\vert 0 \rangle$ 
define a permutation $P$ of those, $\{ i_k \}$, of the state 
$\vert a \rangle$. As is well known, one can 
always resolve the total permutation $P$ as a product of NN permutations 
$ P=\prod_{k=1}^Q P_{l_k,l_k+1} $, i.e., 
$\vert b \rangle =  P \vert a \rangle$. 
The Hamiltonian (\ref{tJ}) is the 
sum $H=-\sum_{i=1}^N {\cal P}_{i,i+1}$ of NN operators 
${\cal P}_{i,i+1}= t\,P^K_{i,i+1}+(J/2)\,P^{M}_{i,i+1}$. 
By definition, the expansion of $H^Q$ contains the product 
${\cal P} = \prod_{k=1}^Q {\cal P}_{l_k,l_k+1}$, where $\{l_k\}$ is 
exactly the sequence of site indices entering the decomposition of $P$. 
If $t$ and $J$ are both non-zero one has 
${\cal P}\vert a \rangle=\nu P \vert a \rangle$, where $\nu$ 
-- if $Q$ is assumed as the {\em minimum} integer allowing the 
decomposition -- is a strictly positive constant. Hence, we obtain 
$\vert \langle b \vert H^Q\vert a\rangle \vert \ge  \nu$, because 
all other sequences of permutations contained in the expansion of $H^Q$ 
that in case link the states $\vert a\rangle$ and $\vert b\rangle$  
contribute to the expectation value of 
$\langle b \vert H^Q\vert a\rangle$ with the same sign. 
This concludes the proof of non-degeneracy of the 
ground-state for $t\ne 0$ and $J\ne 0$.

$\bullet$ The uniqueness of the ground-state (henceforth we avoid the case 
$t=0$ or $J=0$) is valid in any sector with fixed colour numbers, but of 
course one can have many degeneracies among distinct sectors. Indeed, the 
symmetries (\ref{comm}) allow a simple proof that:
\begin{description}
\item[(C)] {For an open chain, the ground-state energies
of the $B^xF^y_J$ and of the $BF^y_J$ system are degenerate}.
\end{description}
In fact, from property (A) the ground-state $\vert \psi_0 \rangle$ 
of the $B^xF^y_J$ system in the
$N^{y+1}=N^{y+2}=...=N_{z-1}=0$, $N^z=N_B$ sector is the ground-state of 
the $B F^y_J$ system. This state is a maximal weight for the algebra 
SU$(x)$ annihilated by all {\em bosonic} raising operators, 
${\Gamma^{z}}_{\alpha}\vert \psi_0\rangle=0$, for $\alpha<z$, 
$\xi_{\alpha}=1$, so that by applying on it 
bosonic lowering operators ${\Gamma^{\alpha}}_z$ one can obtain a degenerate 
eigenstate of any sector of the $B^xF^y_J$ system with arbitrary bosonic 
occupation numbers $N^\alpha$ but fixed $N_B$. These kind of states satisfy 
the condition $\gamma_a=\vert \gamma_a\vert$, and from the uniqueness 
property we conclude that each of them must be the ground-state of its 
sector. This concludes the proof. 
By taking the thermodynamic limit one can then easily show that the 
energy per site does not depend on the boundary conditions\cite{yang}, 
confirming and extending to the general $J\ne 2\,t$ case the Sutherland 
conclusion\cite{suth} that the $B^x F^y$ and the  $B F^y$ systems have not 
only the same ground-state energy per site but also the {\em same }
ground-state (both equivalent by the SU$(x)$ symmetry). 
This statement however holds only in one dimension, and cannot be extended 
to higher dimensions, where counter-examples exist\cite{as}.
 
$\bullet$ The uniqueness of the ground-state $\vert \psi \rangle$ 
also implies that all its quantum numbers are well defined. As usual in 
this case, these numbers can be identified by selecting some simple reference 
state $\vert \phi \rangle=\sum_a\rho_a\vert a \rangle$ (e.g., 
ground state of a simple reference Hamiltonian) with non negative  
amplitudes $\rho_a = \vert \rho_a\vert $ and for which the quantum numbers 
are known.  By the non vanishing value of the scalar product 
$\langle \phi\vert \psi\rangle = \sum_a \vert \rho_a \gamma_a \vert$ 
it follows that the ground-state must belong to the same sector. 

The discussion of the ordering of energy levels and their associated 
quantum numbers in its generality is quite involved. For the sake of 
clarity, we now specialize to the simpler but physically most interesting 
case of the $B^2F^2_J$ system. For convenience we rename the 
Hubbard operators 
\begin{eqnarray}
&& T_{+,i} = \Gamma^1_{i,2},\quad 
   T_{-,i} = \Gamma^2_{i,1},\quad 
   T_{z,i} = {1\over2}   (\Gamma^1_{i,1} - \Gamma^2_{i,2}), 
\label{spinT} \\
&& L_{+,i} = \Gamma^3_{i,4},\quad 
   L_{-,i} = \Gamma^4_{i,3},\quad 
   L_{z,i} = {1\over2}   (\Gamma^3_{i,3} - \Gamma^4_{i,4}),
\label{spinL}
\end{eqnarray}
and denote ${\vec T_i}= (T_{x,i}, T_{y,i} ,T_{z,i})$, where
$T_{\pm,i}=T_{x,i} \pm i T_{y,i}$, and similarly for ${\vec L_i}$. 
We also denote ${\vec T} =\sum_{i=1}^N {\vec T_i}$ and  
${\vec L} =\sum_{i=1}^N {\vec L_i}$, which thus are the fermion and boson 
total spin operators, respectively. The eigenstates are classified 
according to the quantum numbers $N_F$, $T_z$, $L_z$, $T$, $L$, 
where $T(T+1)$ and $L(L+1)$ are the eigenvalues of 
$T^2 = {\vec T}\cdot{\vec T}$ and $L^2 = {\vec L}\cdot{\vec L}$, 
respectively. 

Result (C) and Eq.~(\ref{spinL}) imply that the ground-state of the 
$B^2F^2_J$ system is characterized by maximum value of the boson total 
spin, $L=\vert L_z \vert=N_B/2$. To discuss the value of the fermion 
total spin in the ground-state, we follow the approach of Lieb and 
Mattis\cite{lieb-mattis} for the case of the Heisenberg model. 
After division of the chain into two sublattices $A$ and $B$, we define 
the reference Hamiltonian 
\begin{equation}
{\tilde H} = 
2\, {\vec T}_A\cdot {\vec T}_B , 
\label{HRef}
\end{equation} 
where ${\vec T}_{p} = \sum_{i\in p} {\vec T}_i$ ($p=A,B$). 
The Hamiltonian (\ref{HRef}) commutes with local fermion 
$n_i^f=\Gamma^1_{i,1}+\Gamma^2_{i,2}$ and boson 
$n_i^b=\Gamma^3_{i,3}+\Gamma^4_{i,4}$ occupation numbers. 
Denoting $n^f_p=\sum_{i\in p} n_i^f$ the number of fermions on sublattice 
$p$, the {\em lowest} eigenvalue ${\tilde E}(T,L)$ of ${\tilde H}$ for 
each given $T$ and arbitrary $L$ is easily found
\begin{equation} 
{\tilde E}(T,L) = T(T+1) - {1\over4}\sum_{p=A,B}n^f_p(n^f_p+2) , 
\label{eigenvalue}
\end{equation}
where the allowed values of $T$ are 
\begin{equation}
    {{\vert n^f_A - n^f_B \vert}\over2} \le T \le {N_F\over2}. 
\end{equation}
In the basis (\ref{state}) the sign of the off-diagonal matrix elements 
of Eq.~(\ref{HRef}) is not definite. However, to select a good reference 
state it is sufficient to consider the subset ${\cal G}$  
of the basis vectors where the fermions occupy the 
first $N_F$ sites of the chain. In fact, it is not difficult to check that
{\em in the invariant subspace} ${\cal G}$ the off-diagonal 
matrix elements of ${\tilde H}$ are non-positive,
because in Eq.~(\ref{HRef}) permutations 
take place only between sites of different sublattices. 
The previous analysis ({\ref{perron}) implies therefore  
that  the amplitudes $\rho_a$  of the lowest energy 
eigenstate can be chosen 
non-negative. We conclude that the ground-state of $H$ has the same quantum 
numbers of the reference state. In the subspace ${\cal G}$ one has 
\begin{eqnarray}
n^f_A={{N_F}\over2},\,\,\,\,n^f_B={{N_F}\over2}\quad\quad\qquad &
{\rm for} & \quad N_F\quad{\rm even}, \nonumber \\
n^f_A={{N_F \pm 1 }\over2},\,\,
n^f_B={{N_F\mp 1}\over2}\quad &
{\rm for}& \quad N_F\quad{\rm odd}, 
\end{eqnarray}
and by minimizing Eq.~(\ref{eigenvalue}) one finds that 
the ground-state of the $B^2F^2_J$ system has $T=0$ (for $N_F$ even) 
or $T=1/2$ (for $N_F$ odd). More in general, following 
Ref.~{\cite{lieb-mattis}, from Eq.~(\ref{eigenvalue}) we can order 
the energy levels $E(T,L)$ of the $B^2F^2_J$ system in function 
of the fermion total spin quantum number 
\begin{equation}
E(T,L)<E(T+1,L). 
\label{levels}
\end{equation}

$\bullet$ Eq.~(\ref{levels}) and the result (A) allow 
the determination of the ordering of the energy levels $E_J$ for the 
antiferromagnetic $t$-$J$ Hamiltonian. 

The correspondence between the $BF^2_J$ system and the $t$-$J$ model for 
$J>0$ is explicitly derived upon identification of 
spin-up and spin-down electrons with the two fermion colours and 
of the hole with  the boson colour $\alpha=3$. Accordingly, we 
have $\vec T\equiv\vec S$ and $N_B\equiv N_h$, where 
$\vec S$ and $N_h$ are the physical spin and hole number, respectively. 
Because $BF^2_J$ is the subspace of $B^2F^2_J$ with 
$L=\vert L_z\vert=N_B/2$, from Eq.~(\ref{levels}) we derive the 
ordering $E_{J>0}(S,N_h)<E_{J>0}(S+1,N_h)$. In particular, we find that 
for $J>0$ the ground-state of the $t$-$J$ model on an open 
chain {\em has minimum total spin $S$ and is non degenerate} (modulo 
the trivial twofold $S_z=\pm1/2$ spin degeneracy for $N_h$ odd). 

In a similar fashion, the correspondence between the $B^2F_J$ and the 
$t$-$J$ model for $J<0$ is obtained by identifying spin-up and spin-down 
electrons with the two boson colours and the hole with the fermion 
colour $\alpha=1$. In this case we have $\vec L\equiv\vec S$ and 
$N_F\equiv N_h$, so that by applying result (C) we find that for 
$J<0$ the ground-state of the $t$-$J$ model on an open chain is [modulo the 
trivial $(N-N_h+1)$-fold spin degeneracy] a {\em nondegenerate fully 
polarized ferromagnet}. 

Concerning the general case, we emphasize here that the degeneracy among 
sectors with different number of bosonic colours but with the same $N_B$ 
is due to the fact that the quadratic operator
\begin{equation}
{\cal C}_2^x =  \sum_{\alpha,\beta>y} 
{\Gamma^{\alpha}}_{\beta}{\Gamma^{\beta}}_{\alpha}
\label{casimir}
\end{equation}
of the bosonic algebra assumes its maximum value in the  ground-state.   
Concerning the fermionic SU$(y)$ sector for generic $y>2$, symmetry 
considerations, the case $y=2$, as well as the exact solution at the 
supersymmetric point\cite{suth}, suggest that the ground-state 
of the Hamiltonian (\ref{tJ}) should have {\em minimum} value of the 
quadratic operator ${\cal C}_2^y$ [namely, the operator as defined in 
Eq.~(\ref{casimir}) but for fermions, i.e., $\alpha,\beta\le y$]. 
In particular, for values of the fermion number $N_F=yn$, with $n$ an 
integer, we conjecture that the absolute ground-state of the multicomponent 
model (\ref{tJ}) {\em necessarily} belongs to the sector with 
$N^1=...=N^y=n$ for any value of the coupling $J>0$. 

We have presented a multicomponent generalization of the $t$-$J$ model, 
extending in a natural way the well known Sutherland's Hamiltonian
corresponding to the supersymmetric $J=\pm 2\,t$ cases.
In one dimension we have determined that doping does not alter the 
 ground-state spin sector of the Heisenberg Hamiltonian, nor the 
ordering of energy levels of the antiferromagnetic model. 
These statements have been derived by using Perron-Frobenius 
theorem, i.e., by exploiting the irrelevance between fermion and boson 
statistics, typical for one dimensional systems with NN interactions. 
Indeed, the off-diagonal matrix elements of the Hamiltonian (\ref{tJ}) 
can be made non-positive by a proper choice of the fermion-boson basis.
This property is clearly lacking in higher dimensions, and consequently 
the equivalence between the $B^x F^y_J$ and the $B F^y_J$ systems, valid 
in one dimension, is {\em not rigorous } in higher dimensions -- unless 
for the trivial case without fermions. In fact, contrary to what 
previously stated\cite{suth,EKS}, it is easy to find a counter-example, 
where the ground-state of the $B^2 F_J$ system (i.e., the ferromagnetic 
$t$-$J$ model) does not belong to the sector of maximum spin $B F_J$, 
(i.e., to the Nagaoka sector) but to the singlet sector\cite{as}.

Both Authors acknowledge useful discussions with C. Castellani, E. Tosatti,
and F. E{\ss}ler, and are grateful to I.S.I. for the kind hospitality in 
Torino, I.S.I. Contract ERBCHRX-CT920020. A.A. also acknowledge the 
University of Udine for warm hospitality.


\begin{references}
\bibitem{susy} At the supersymmetric point one has the strongest 
               condition $[H_S,{\Gamma^{\alpha}}_{\beta}]=0$ 
               $\forall \alpha,\beta$. Because ${\Gamma^{\alpha}}_{\beta}$ 
               for $\xi_{\alpha}\ne\xi_{\beta}$ necessarily 
               changes $N_F$ ($N_B$) by $\pm1$, for a $B^xF^y$ system there 
               are additional degeneracies among states with different 
               fermion number $N_F$ and the degenerate multiplets belong to
               the Spl$(x\vert y)$ supergroup.
\end{references}
\end{document}